\documentclass[preprint,aps,eqsecnum]{revtex4}
\usepackage{epsfig}
\begin{document}
\newcommand{\be}{\begin{eqnarray}}
\newcommand{\ee}{\end{eqnarray}}

\preprint{DO-TH 03/09}
\vspace*{1cm}
\title
{Manifestly Covariant Analysis of the QED Compton Process in {\mbox
{\boldmath $e p\rightarrow e \gamma p$}} and {\mbox{\boldmath $e p 
\rightarrow e \gamma X $ }}}
\author{\bf A. Mukherjee}
\email{asmita@physik.uni-dortmund.de}
\author{C. Pisano}\email{pisano@physik.uni-dortmund.de}   
\affiliation{ Institut f\"ur Physik, Universit\"at Dortmund, D 44221
Dortmund, Germany}

\date{\today\\[2cm]}

\begin{abstract}

We calculate the unpolarized QED Compton scattering cross section in a 
manifestly  covariant way. Our approach  allows a
direct implementation of the specific kinematical cuts imposed in the 
experiments, {\it e. g.} HERA-H1.
We compare the 'exact' cross section in terms of the structure functions
$F_{1,2} (x_B,Q^2)$, assuming the Callan-Gross relation, with the one  obtained
using the equivalent photon approximation (EPA) as well as with the experimental
results. We find that the agreement with the
EPA is better in $x_{\gamma}$ bins, where $x_{\gamma}$ is the fraction of
the longitudinal momentum of the proton carried by the virtual photon,
compared to the  bins in the leptonic variable $x_l$.  
\end{abstract}
\maketitle

\section{Introduction}

The QED Compton scattering in high energy electron-proton collisions 
 $ e(l)p(P)\rightarrow e(l')\gamma(k')X$,
with a real photon $\gamma(k')$ emitted at the lepton vertex (Fig.1), is
one of the most important processes for an understanding of the photon
content of the proton in the so-called 
'equivalent photon' approximation (EPA), first introduced for a charged 
particle 
by Weizs\"acker and
Williams \cite{ww} and later extended to the nucleon and investigated
widely \cite{kniehl,drees,ohn,gsv,ruju,gpr1,gpr2}. QED Compton scattering has
been recently analyzed by the H1 collaboration at HERA \cite{lend}.
In addition to the information concerning the photon content of the
proton within the framework of EPA, 
it can also shed some light on the structure functions $F_{1,2}(x_B,Q^2)$ 
of the proton  \cite{lend,kessler}  in the low-$Q^2$
region, which are presently poorly known \cite{blu}. In \cite{lend} these 
alternative
descriptions were confronted with data and it was found that the description
in terms of  $F_{1,2}$, {\it i.e.} for $X \not= p$, is superior to the one
in terms of the inelastic photon distribution $\gamma_{\mathrm{inel}} (x_B,Q^2)$.
Henceforth we shall refer to the description in terms of $F_{1,2}$ as
'exact' to distinguish it  from the approximations involved in the EPA. It
should be noted, however,  that 
the analysis in \cite{lend,kessler} utilized the Callan-Gross
relation \cite{callan}
$ F_L(x_B,Q^2)=F_2(x_B,Q^2)-2 x_B F_1 (x_B,Q^2)=0$. This relation is
contaminated by higher order (NLO) QCD corrections as well as by higher
twist contributions relevant in the low-$Q^2$ region which may invalidate     
the assumptions underlying the 'exact' analysis.
Furthermore, the analysis in \cite{lend,kessler} was carried out
within the framework of the helicity amplitude  formalism \cite{kessler}.
The implementation of experimental cuts within this formalism is nontrivial
and affords therefore an iterative numerical approximation procedure
\cite{kessler,h1} whose first step  corresponds to $-k^2=Q^2=0$, where $k$
is the momentum of the virtual photon.

It is this second issue which we intend to study here. We shall replace the
noncovariant helicity amplitude analysis of \cite{kessler} by a standard
covariant tensor analysis whose main advantage, besides compactness and
transparency, is the possibility to implement the experimental cuts directly
and thus avoid the necessity of employing an iterative approximation of
limited accuracy. The first issue concerning the $F_L$ contributions affords
some estimates of this poorly known structure function and we refrain  from
its study here.

The plan of the paper is as follows. In section II, we calculate the exact 
cross section  for the elastic scattering. In section III, we calculate the
cross section for the inelastic channel. Our numerical results are discussed in
section IV. The summary is given in section V. All the useful and necessary formulae and
the kinematics are given in the appendices A, B, C and D.

\section{Elastic QED Compton Scattering}
We consider elastic QED Compton scattering:
\be
e(l)+p(P) \rightarrow e(l')+\gamma(k')+ p(P'),
\ee
where the 4-momenta of the particles are given in the brackets. 
We introduce the invariants 
\be
S=(P+l)^2, ~~~~~~~~~\hat s=(l+k)^2, ~~~~~~~~t=k^2.
\label{invar}
\ee
Here $k=P-P'$ is the 4-momentum of the virtual photon. 
The photon in the final state is real, $k'^2=0$. We neglect the
electron mass everywhere except when it is necessary to avoid divergences in
the formulae and take the proton to be massive, $ P^2=P'^2=m^2 $. The
relevant Feynman diagrams for this process are shown in Fig. 1, with $X$
being a proton and $P_X=P'$. The
squared matrix element can be written as
\be
\overline{{\mid M_{\mathrm{el}} \mid }^2}={1\over t^2} H^{\mu \nu}_
{\mathrm{el}}(P,P') 
T_{\mu \nu}(l,k;l',k'),
\ee
where
\be
H^{\mu \nu}_{\mathrm{el}}(P,P') = {1\over 2}\sum_{\mathrm{{spins}}} {\langle 
P' \mid J^\nu 
(0)\mid P \rangle}^* \langle P' \mid J^\mu (0) \mid P \rangle
\label{hmunu}
\ee
is the hadronic tensor, $J^\mu$ being the electromagnetic current.

If we use the notation 
\be
dPS_N(p;p_1,...,p_N)=(2 \pi)^4 \delta(p-\sum_{i=1}^N p_i) \prod_{i=1}^N
{d^3p_i\over { (2 \pi)^3 2 p_i^0}}
\ee
for the Lorentz invariant $N$-particle phase-space element, the total 
cross section will be
\be
\sigma_{\mathrm{el}}(S)={1\over {2 (S-m^2)}} \int dPS_{2+1}(l+P;l',k',P')
\overline{{\mid M_{\mathrm{el}} \mid }^2}~.
\label{sigmael}
\ee
Eq. (\ref{sigmael}) can be rewritten following the technique of \cite{kniehl},
which we slightly modify to implement the experimental cuts and 
constraints; in particular all the integrations are  performed
numerically. Rearranging the ($2+1$)-particle phase space into a sequence
of two $2$-particle ones, Eq. (\ref{sigmael}) becomes:
\be
\sigma_{\mathrm{el}}(S)={1\over {2 (S-m^2)}} \int {d\hat s\over 2 \pi} \,dPS_2
(l+P;l'+k',P') {1\over t^2} H^{\mu \nu}_{\mathrm{el}} (P,P') X_{\mu \nu}(l,k)~.
\ee
$X_{\mu \nu}$ contains all the informations about the leptonic part of the
process and is defined as
\be
X_{\mu \nu}(l,k)=\int dPS_2(l+k;l',k') T_{\mu \nu}(l,k;l',k'),
\ee
$T_{\mu \nu}$ being the leptonic tensor \cite{ji,anf}:
\be
T_{\mu \nu}(l,k;l',k') =\,{{4 e^4}\over {\hat{s}\hat{u}}}\,\bigg
\{ \,{{1}\over{2}}\,
g_{\mu\nu}\,(\hat s^2 + \hat u^2 + 2 \hat t t) + 2\hat s\, l_{\mu}l_{\nu}
+ 2 \hat u\,l_{\mu}'l_{\nu}'~~~~~~~~~~~~~~~~~~\nonumber \\
~~~+\,(\hat t+t)(l_{\mu}l_{\nu}'+l_{\nu} l_{\mu}') 
-({\hat s}-t)\,(l_{\mu}k_{\nu}'+l_{\nu}k_{\mu}')~~~~~~~~\nonumber \\ 
+\,(\hat u -t) \, (l_{\mu}'k_{\nu}' +l_{\nu}'k_{\mu}')\bigg\},
\label{lept}    
\ee
where we have defined $\hat t=(l-l')^2$ and $\hat u = (l-k')^2$.
It can be shown that
\be
dPS_2(l+k;l',k')={d\hat t\, d\varphi^*\over {16 \,\pi^2 (\hat s-t)}},
\ee
with  $\varphi^*$ denoting the azimuthal angle of the outgoing $e-\gamma$ 
system in the $e-\gamma$ c.m. frame.
For unpolarized scattering, $X_{\mu \nu}$ is symmetric in the indices
$\mu$, $\nu$  and can be expressed in terms of the two 
Lorentz scalars $\tilde{X}_1$ and $\tilde{X}_2$:
\be
X_{\mu \nu}(l,k)&=&{1\over 2 t} \bigg \{ [ 3\tilde{X}_1(\hat s,t
)+\tilde{X}_2 (\hat s,t) ] \bigg
({2 t\over \hat s-t} l-k \bigg )_\mu \bigg ({2 t\over \hat s-t} l-k\bigg )_\nu
\nonumber \\&&~~~~~+[ \tilde{X}_1(\hat s,t)+\tilde{X}_2(\hat s,t)] (t g_{\mu \nu}-k_\mu k_\nu)
\bigg \},
\label{xmunu}
\ee
with
\be
\tilde{X}_1(\hat s,t)= {4 t\over (\hat s-t)^2} l^\mu l^\nu X_{\mu \nu}(l,k),
\ee
\be
\tilde{X}_2(\hat s,t)=g^{\mu \nu} X_{\mu \nu}(l,k).
\ee
Using the leptonic tensor (\ref{lept}) and also the relations
\be
l \cdot k= {1\over 2} (\hat s-t), ~~~~l \cdot P={1\over 2} (S-m^2),~~~k
\cdot P={1\over 2} t,
\ee
we obtain
\be
{t\, l^\mu l^\nu T_{\mu \nu}\over {4 \pi^2 (\hat s-t)^3}}=\,
{e^4}\, {-t \hat t\over {2 \pi^2 (\hat s-t)^3}} \equiv X_1(\hat s,t,\hat t),
\label{x1}
\ee
\be
{g^{\mu \nu} T_{\mu \nu}\over {16 \pi^2 (\hat s-t)}}= \,{e^4}\, {(t^2-2 t
\hat s+2 \hat s^2+2 \hat s \hat t+\hat t^2)\over {4 \pi^2 \hat s (\hat s-t)
(t-\hat s-\hat t)}} \equiv X_2(\hat s,t, \hat t),
\label{x2}
\ee
where $e^2=4 \pi \alpha $.
The invariants $X_i(\hat s,t,\hat t)$, with $i=1,2$, are related to $\tilde{X}
_i(\hat s,t)$ by
\be
\tilde{X}_i(\hat s,t)=2 \pi \int_{\hat t_{\mathrm{min}}}^
{\hat t_{\mathrm{max}}} d \hat t\,  {X}_i(\hat s,t,\hat t).
\ee
The integration limits of $\hat t$ are:
\be
\hat t_{\mathrm{max}}=0, ~~~~\hat t_{\mathrm{min}} = -\hat s +t+{\hat s  
\over \hat s-t} \, m_e^2,
\label{thatlim}
\ee
where $m_e$ is the mass of the electron. We point out that the kinematical cuts 
employed by us prevent the electron propagators to become too small and 
thus the divergences are avoided, so we can safely
neglect the electron mass in the numerial calculation. 
The hadronic tensor in the case of elastic scattering can be expressed in 
terms of the common proton form factors as
\be
H^{\mu \nu}_{\mathrm{el}}(P,P')=  e^2\,[ H_1(t) (2 P-k)^\mu (2 P-k)^\nu +
H_2(t) (t g^{\mu
\nu}-k^\mu k^\nu)],
\ee
with
\be
H_1(t)={{G_E^2(t)- (t/4 \,m^2)\, G_M^2(t)}\over{1-{t/4\, m^2}}},~~~~~~~~ H_2(t)=G_M^2(t).
\ee
The electric and magnetic form factors are empirically parametrized
as dipoles: 
\be
G_E(t)= {1\over [1 - t/(0.71\,\mathrm{GeV}^2)]^{2}},~~~~G_M(t)=2.79~G_E(t).
\ee
Using 
\be
dPS_2(l+P;l'+k',P')={dt\over 8 \pi (S-m^2)}, 
\ee
finally we get
\be
\sigma_{\mathrm{el}}(S)&=&{\alpha\over 8 \pi (S-m^2)^2} \int_{\hat s_{\mathrm
{min}}}^{(\sqrt
S-m)^2} d\hat s \int_{t_{\mathrm{min}}}^{t_{\mathrm{max}}}{dt \over t} \int_
{\hat t_{\mathrm{min}}}^{\hat t_{{\mathrm{max}}}} d \hat t \int_0^{2 \pi} d \varphi
^* \bigg\{ \bigg [ 2\, {S-m^2\over \hat s-t}
\bigg ( {S-m^2\over \hat s-t}-1 \bigg ) \nonumber\\&& ~~~\times [3 X_1(\hat s,t,\hat
t)+X_2 (\hat s,t,\hat t)] +{2 m^2\over t} [X_1(\hat s,t,\hat t)+X_2(\hat
s,t,\hat t)]+X_1(\hat s,t,\hat t)\bigg ]
H_1(t)\nonumber\\&&~~~~~~~~~~~~~~~~~~~~+X_2(\hat s,t,\hat t) H_2(t) \bigg \},
\label{sigel}
\ee
where $\hat s_{\mathrm{min}}$ denotes the minimum of $\hat s$ and $t_{\mathrm
{min},\, {\mathrm{max}}}$ are given by
\be
t_{\mathrm{min},\,{\mathrm{max}}}=2 m^2-{1\over 2  S} \Big [ (S+m^2)  (S-\hat 
s+m^2) \pm (S-m^2) \sqrt
{(S-\hat s+m^2)^2-4 S m^2}\,\Big ].\nonumber \\
\ee 
It is to be noted that in Eq. (\ref{sigel}) we have shown the integration
over $\varphi^*$ explicitly, because of the cuts that we shall impose on the 
integration variables for the numerical calculation of the cross section. 
The cuts are discussed in section IV.  
The EPA consists of considering the exchanged
photon as real, so it is particularly good for the elastic process in which
the virtuality of the photon $|t|$ is constrained to be small ($\lesssim 1\, 
\mathrm{GeV}^2$) by the form factors. It is possible to get the approximated
cross section $\sigma_{\mathrm{el}}^{\mathrm{EPA}}$ from the exact one in a 
straightforward way, following again {\cite{kniehl}}. If the invariant mass
of the system $e-\gamma$ is large compared to the proton mass, $\hat s_{\mathrm
{min}} \gg m^2$, one can neglect $ |t|  $ versus $\hat s $, $m^2$ versus $S$, 
then
\be
{X}_1(\hat s, t,\hat t) \approx {X}_1(\hat s,0, \hat t)=0,
\label{xone}
\ee
and \be
{X}_2(\hat s, t, \hat t)\approx {X}_2(\hat s,0, \hat t) = -{{2 \hat s}\over{\pi}}\, {{d\hat{\sigma}(\hat s, \hat t)}\over{d\hat t}},
\label{xtwo}
\ee
where we have introduced the differential cross-section for the real 
photoproduction process $e \gamma \rightarrow e \gamma$:
\be
{d \hat \sigma (\hat s, \hat t)\over d\hat t}=-{2 \pi \alpha^2\over {\hat
s}^2}
\Bigg ( 
{\hat u\over \hat s}+{\hat s\over \hat u} \Bigg ),
\ee
with $\hat u = -\hat s -\hat t$.
We get:
\be
\sigma_{\mathrm{el}}(S) \approx \sigma_{\mathrm{el}}^{\mathrm{EPA}} =
\int_{x_{\mathrm{min}}}^{(1-{m/ 
\sqrt S})^2}\, dx \,\int_{m_e^2 -\hat s}^0
d\hat t  \,\gamma_{\mathrm{el}} (x) \,{d \hat \sigma (x S, \hat t)\over d\hat t} ,
\label{epael}
\ee
where $x={\hat s/ S}$ and $\gamma_{\mathrm{el}}(x)$ is the elastic
contribution to the equivalent 
photon distribution of the proton \cite{kniehl,gpr1}:
\be
\gamma_{\mathrm{el}}(x) =-{\alpha\over 2 \pi} x \int_{t_{\mathrm{min}}}^{t_{\mathrm
{max}}} {dt\over t} \bigg \{ 2 \bigg [ {1 \over x}\bigg  ( {1\over x}-1 \bigg ) 
+{m^2\over t}\bigg ] H_1(t) + H_2(t) \bigg \},
\ee
with
\be
{t_{\mathrm{min}} \approx -\infty} ~~~~~~~~~~~~t_{\mathrm{max}} \approx 
-{{m^2 x^2}\over{1-x}}.
\ee
To clarify the physical meaning of $x$, let us introduce the variable $x_{\gamma}$:
\be
x_{\gamma} = {{l \cdot k} \over {P \cdot l}}.
\ee
It is possible to show that $x_{\gamma}$ represents the fraction  of the 
longitudinal momentum of the proton carried by the virtual photon, 
so that one can write
\be
k = x_{\gamma} P + \hat k,
\ee   
with $\hat k \cdot P = 0$. Using (\ref{invar}) one gets
\be
x_{\gamma} = {{\hat s - t}\over{S-m^2}},
\label{xgamma}
\ee
which reduces to $x$ in the EPA limit.  One can also 
define the leptonic variable $x_l$:
\be
x_l = {{Q^2_l}\over{2 P \cdot (l - l')}},
\label{xl}
\ee
where $Q^2_l = - \hat t$. When $t \simeq 0$, it turns out that also
$x_l \simeq x$.


\section{Inelastic QED Compton Scattering}
To calculate the inelastic QED Compton scattering cross section, we extend the
approach discussed in the previous section. In this case, an electron and a
photon are produced in the final state with a general hadronic system $X$.  
In other words, we consider the process
\be
e(l)+p(P) \rightarrow e(l')+\gamma(k')+X(P_{X}),
\ee
where $P_{X}=\sum_{X_i} P_{X_i}$ is the sum over all momenta of the produced 
hadronic system.
Let the invariant mass of the produced hadronic state $X$ to be $W$. Eq. (\ref
{invar}) still holds with $Q^2 = -t$.
The cross section for inelastic scattering will be
\be
\sigma_{\mathrm{inel}}(S)={1\over {2 (S-m^2)}} \int dPS_{2+N}(l+P;l',k',P_{X_1}, 
..., P_{X_N})
\overline{{\mid M_{\mathrm{inel}} \mid }^2},
\label{sigmainel}
\ee
where 
\be
\overline{{\mid M_{\mathrm{inel}} \mid }^2}={1\over Q^4} H^{\mu \nu}_
{\mathrm{inel}}(P,P_X) 
T_{\mu \nu}(l,k;l',k')
\ee
is the squared matrix element and
\be
H^{\mu \nu}_{\mathrm{inel}}(P,P_X) = {1\over 2}\sum_{\mathrm{{spins}}} \sum_{X} 
{\langle P_X \mid J^\nu 
(0)\mid P \rangle}^* \langle P_X \mid J^\mu (0) \mid P \rangle~.
\ee
If we rearrange the ($2+N$)-particle space phase into a sequence of a $2$-particle and a $N$-particle one, we get  
\be
\sigma_{\mathrm{inel}}(S)={1\over 2 (S-m^2)} \int {d W^2\over 2 \pi} \int {d 
\hat s\over 2 \pi } \int dPS_2(l+P;l'+k',P_{X}) {1\over Q^4} W^{\mu \nu}
(P,k) 
X_{\mu \nu} (l,k),\nonumber \\
\ee
where $X_{\mu \nu}$ is given by Eq. (\ref{xmunu}) and $W^{\mu \nu}$ is the
hadronic tensor for inelastic scattering
\be
W^{\mu \nu}= \int dPS_{N}(P-k;P_{X_1},....,P_{X_N})\,H^{\mu \nu}_{\mathrm{inel}}~.
\ee 
The hadronic tensor is parametrized as
\be
W^{\mu \nu}={4 \pi e^2 \over Q^2} \bigg [- (Q^2 g^{\mu \nu} + k^{\mu} k^{\nu}) 
F_1(x_B,Q^2)+(2x_B P^{\mu}-k^{\mu}) (2 x_B P^{\nu}-k^{\nu}){F_2(x_B,Q^2)\over 
2 x_B} \bigg ], \nonumber \\
\ee
where $x_B$ is the Bjorken variable given by
\be
x_B= -{Q^2\over 2 P\cdot k} = {Q^2\over Q^2+W^2-m^2}.
\ee
Using 
\be
dPS_2(l+P;l'+k',P_X)= {dQ^2\over 8 \pi (S-m^2)} 
\ee
as before, we get
\be
\sigma_{\mathrm{inel}}(S)&=&{\alpha\over 4 \pi (S-m^2)^2} \int_{W^2_{{\mathrm
{min}}}}^{W^2_{\mathrm{max}}}
 d W^2 \int_{\hat s_{\mathrm{min}}}^{(\sqrt
S-W)^2} d\hat s \int_{Q^2_{\mathrm{min}}}^{Q^2_{\mathrm{max}}} {dQ^2 \over 
Q^4} \int_{\hat t_{\mathrm{min}}}^{\hat
t_{max}} d \hat t \int_0^{2 \pi} d \varphi^* 
\bigg \{ \bigg [\bigg ( 2 \,{{S-m^2}\over {\hat s+Q^2}}\nonumber\\&&~
\times \bigg (1-{{S-m^2}\over {\hat s+Q^2}}\bigg )+(W^2-m^2) \bigg ( {2\,
(S-m^2)\over {Q^2 (\hat s + Q^2)}}-{1\over Q^2} +{m^2-W^2\over 2 \,Q^4}\bigg ) 
\bigg )
\nonumber\\&&~~~~\times [3
X_1(\hat s,Q^2,\hat t)+X_2(\hat s,Q^2,\hat t)]+\bigg ({1\over
Q^2}(W^2-m^2)+{(W^2-m^2)^2\over 2\, Q^4}+{2 m^2\over Q^2} \bigg  )
\nonumber\\&&~~~~~~~\times [X_1(\hat s,Q^2,\hat t)+X_2(\hat s,Q^2,\hat t)]-X_1(\hat
s,Q^2,\hat t)\bigg ] F_2(x_B,Q^2) {x_B\over
2}\nonumber\\&&~~~~~~~~~~~~~~~~~~~~~~~~~~~~~~~~~~-X_2(\hat s,Q^2,\hat
t)F_1(x_B,Q^2)\bigg
\}.
\label{siin}
\ee 
Here $X_i(\hat s,Q^2,\hat t)$, with $\,i=1,2 $, are given by Eqs. (\ref{x1})-(\ref{x2}) with $t$ replaced by $-Q^2$. The limits of the integration over
$Q^2$ are:
\be
Q^2_{{{\mathrm{min}},{\mathrm{max}}}}=-m^2-W^2+{1\over 2 S} \Big [(S+m^2) (S-\hat s+W^2) \mp (S-m^2) {\sqrt
{(S-\hat s+W^2)^2-4 S W^2}} \Big ],\nonumber\\
\ee  
while the extrema of $\hat t$ are the same as Eq. (\ref{thatlim}).
The limits $W^2_{\mathrm{min},\mathrm{max}}$ are given by:
\be
W^2_{\mathrm{min}}=(m+m_{\pi})^2,~~~~~~~~W^2_{\mathrm{max}}=
(\sqrt S-\sqrt {\hat s_{\mathrm{min}}}\, )^2,
\ee
where $m_{\pi}$ is the mass of the pion. $F_1$ and  $F_2$ are the usual structure
functions of the proton. 

In the EPA, we neglect $m^2$ compared to $S$
and $Q^2$ compared to $\hat s$ as before. Using Eqs. (\ref{xone}) and ({\ref{xtwo}),
 we can write
\be
\sigma_{\mathrm{inel}}(S) \approx \sigma_{\mathrm{inel}}^{\mathrm{EPA}} =
\int_{x_{\mathrm{min}}}^{(1-m/\sqrt S)^2} dx \, \int_{m_e^2 -\hat s}^0
d\hat t~
\gamma_{\mathrm{inel}}(x, x S) \,{{d\hat\sigma(x S, \hat t)}\over{d\hat t}},
\label{epain}
\ee
where again $x={\hat s/S}$  and $\gamma_{\mathrm{inel}}(x, x S)$ is the 
inelastic contribution to the equivalent photon distribution of the proton
\cite{anlf}:
\be
\gamma_{\mathrm{inel}} (x, x S)&=&{\alpha\over 2 \pi} \int_x^1\, dy 
\int_{Q^2_{\mathrm{min}}}^{Q^2_{\mathrm{max}}} {dQ^2\over Q^2}\,{y\over x}  
 \bigg [F_2\bigg ({x\over y},Q^2\bigg )\bigg ({{1+(1-y)^2}\over y^2} -
{2 m^2 x^2\over y^2 Q^2} 
\bigg )\nonumber\\&&~~~~~~~~~~~~~~-F_L \bigg({x\over y},Q^2 \bigg) \bigg ].
\label{gammain}
\ee
Following \cite{lend,kessler} we shall use the  LO Callan-Gross relation
\be
F_L(x_B,Q^2)~ =~ F_2(x_B,Q^2)-2 x_B F_1(x_B,Q^2)~=~0
\ee
in our numerical calculations.
The limits of  the $Q^2$ integration can be approximated as
\be
Q^2_{\mathrm{min}}={x^2 m^2\over 1-x}, ~~~~~~Q^2_{\mathrm{max}}= \hat s.
\ee
Our expression of $\gamma_{\mathrm{inel}}(x, xS)$ differs from 
\cite{blu} by a (negligible) term proportional to $m^2$.

\section{Numerical Results}

In this section, we present an estimate of the cross section, calculated
both exactly and using the equivalent photon approximation of the proton.
We have used the same kinematical cuts as used by the H1 collaboration at HERA 
\cite{lend}, which are slightly different from the ones in \cite{kessler}. 
They are imposed on the following lab frame variables: energy of the 
final electron $E_e'$, energy of the final  photon $E_{\gamma}'$, 
polar angles  of the outgoing electron and photon, $ \theta_e$ and 
$\theta_{\gamma}$ respectively,  and  acoplanarity angle $\phi$, 
which is defined as $\phi=|\,\pi-|\phi_{\gamma} -\phi_{e}|\,|$, 
where $\phi_{\gamma}$ and $\phi_{e}$ are the azimuthal angles of 
the outgoing photon and electron respectively ($0 \le \phi_{\gamma},\, 
\phi_e \le 2 \,\pi$). 
The cuts are given by:  
\be
E_e',E_{\gamma}' > 4\, {\mathrm{GeV}}, ~~~~~~~~E_e' + E_{\gamma}'> 20\, 
{\mathrm{GeV}},
\label{cut1}
\ee
\be
0.06 < \theta_e, \theta_{\gamma} <  \pi-0.06,
\label{cut2}
\ee
\be
0 < \phi < {\pi\over 4}.
\label{cut4}
\ee
The energies of the incoming particles are: $E_e = 27.5\,\, 
\mathrm{GeV}$ (electron) and $E_p = 820 \,\,\mathrm{GeV}$ (proton). 
In our conventions, we fix the lab frame such that $\phi_e = 0$, so the 
acoplanarity will be $\phi=|\,\pi-\phi_{\gamma}\,|$.
These cuts reflect experimental acceptance constraints as well as the
reduction of the background events due  to emitted photons with $ (l'+k')^2
\approx 0$ and/or $(l-k')^2 \approx 0 $, which are
unrelated to the QED Compton scattering process (for which $ -k^2=Q^2 
\approx 0$ but
with both $(l'+k')^2$ and $(l-k')^2$ finite), {\it
i.e.} photons emitted parallel to the ingoing (outgoing) electron or from
the hadron vertex \cite{blu,kessler}.
 
We numerically integrate the elastic and inelastic cross sections  
given by  Eqs. (\ref{sigel}) and (\ref{siin}). To implement the cuts in  Eqs.
(\ref{cut1})-(\ref{cut4}), we express $E_e'$, $E_{\gamma}'$, $\cos{
\theta_e}$, $\cos {\theta_{\gamma}}$ and $\cos {\phi}$ in 
terms of our integration variables $ \hat s$, $t$, $\hat t$, $\varphi^*$
(and  $W^2$ in the inelastic channel), as explained in Appendices A-D.  
More explicitly, we use Eqs. (\ref{thetae})-(\ref{phig}), (\ref{cmT}), 
(\ref{cmU}), together with Eqs. (\ref{cmenergy})-(\ref{cmangle2}) for the
elastic channel and Eqs. (\ref{cmenergyin})-(\ref{cmangle2in}) for the 
inelastic one. The cuts imposed on the lab frame variables  
restrict the range of our integrations numerically. In this way, 
we are able to remove  the contributions from  outside the 
considered kinematical region.  

In the calculation of the inelastic cross section, we  have used the ALLM97
parametrization of $F_2(x_B, Q^2)$ \cite{allm}, which is obtained by fitting
DIS data of HERA and fixed target experiments together with the total $p p$
and $\gamma p$ cross sections measured and is expected to hold over the
entire available range of $x_B$ and $Q^2$. We have not considered 
the resonance contribution separately but, using the so called 
local duality \cite{ruju}, we have extended the ALLM97 parametrization of 
$F_2$ from the continuous ($W > 1.8 \,\,\mathrm{GeV}$) down to the 
resonance domain ($m + m_{\pi} < W < 1.8\,\, \mathrm{GeV}$). 
In this way it is possible to agree with the experimental data averaged 
over each resonance, as pointed out in \cite{lend}. The elastic contribution 
to the EPA was calculated using Eq.(\ref{epael}) subject to the additional
kinematical restrictions given by
Eqs. (\ref{cut1}-\ref{cut2}). It corresponds to the same equivalent photon 
distribution as presented
in \cite{ruju,gpr1,gpr2}. For the inelastic channel we
have used Eq. (\ref{epain}) together with Eq. (\ref{gammain}), the cuts
being the same as in the elastic case. We have taken   
$F_L=0$ and used  the ALLM parametrization
of $F_2$, in order to compare consistently with the 'exact' cross section.
We point out that in \cite{ruju,gpr1,gpr2}, $F_2 (x_B, Q^2)$ in  
$\gamma_{\mathrm{inel}}(x,x S)$ was expressed in terms of parton
distributions for which the LO GRV parametrization \cite{grv} was used, 
together with 
$Q^2_{\mathrm{min}}=0.26~ {\mathrm{GeV}}^2 $ so as to guarantee the
applicability of perturbative QCD \cite{gsv}.    
$\gamma_{\mathrm{inel}}(x,x S)$ in our case gives slightly higher results
than the ones obtained with the photon distribution presented 
in \cite{ruju,gpr1,gpr2}.    
 
The Compton process turns out to be dominated by the elastic channel, in fact
after Monte Carlo integration, we find that $\sigma_{\mathrm{el}} = 1.7346$ 
pb, while $\sigma_{\mathrm{inel}} = 1.1719$ pb. The approximated calculation
gives the results: $\sigma_{\mathrm{el}}^{\mathrm{EPA}}= 1.7296$ pb and     
$\sigma_{\mathrm{inel}}^{\mathrm{EPA}}= 1.5969$ pb. This means that in the 
kinematical region under consideration, the total (elastic + inelastic) 
cross section calculated using the EPA agrees with the exact one 
within 14\% and that the approximation  turns out to be particularly good in 
describing the elastic process, for which the agreement is within 0.3\%.  
This is not surprising since in the EPA one
assumes $Q^2 =0$, which is not true especially in the inelastic channel and
the inelastic cross section receives substantial contribution from the
non-zero $Q^2$ region. In terms of the kinematical cuts, the EPA corresponds 
to the situation when the outgoing electron and
the final photon are observed under large polar angles and almost opposite
to each other in azimuth, so that the acoplanarity is approximately zero.
For elastic scattering there is a sharp peak of the exact cross section
for $\phi=0$, contributions from non-zero $\phi$ are very small in this case.
But the inelastic cross section receives contribution even from non-zero
$\phi$, so that in this case the discrepancy from the approximated result is 
 higher. The discrepancy of the total cross section with the approximate
one is thus entirely due to the inelastic part.
       
In Fig. 2  we have compared the total cross sections (exact and EPA) in 
different $x_l $ bins, in the region $1.78 \times 10^{-5} < x_l < 1.78\times
10^{-1}$. Fig. 3 shows that the agreement improves  slightly for bins in 
the variable $x_{\gamma}$. Since $x_{\gamma} \simeq x_l$ for $Q^2\simeq 0$,
the elastic process is not sensitive to this change of variables. 
We point out again that in the EPA limit ($Q^2=0$) $x_l = x_{\gamma} \equiv x$.    
  
In Fig. 4 we show the exact and the EPA cross section in $x_l$ 
and $Q_l^2$ bins together with the experimental results and the 
estimates of the Compton event generator, 
already presented in \cite{lend}.  Except for 
three bins, our exact result agrees with the experiment within the error bars.
The slight difference of our exact result and the one of \cite{lend} 
may be due to the fact that in \cite{lend} the cross section is calculated 
using a Monte Carlo generator in a step by step iteration \cite{kessler,h1} which starts by 
assuming $Q^2=0$, while we did not use any approximation. Our exact result 
is closer to  the EPA in most of the kinematical 
bins as compared to \cite{lend}. The total cross section in the EPA lies above
the 'exact' one in most of the bins.
  
For completeness, we have shown the numerical values of the exact 
and EPA double differential cross sections, 
both for the elastic (Table 1) and inelastic 
(Table 2) contributions. The kinematical bins are the same as in \cite{lend}. 
The exact results when the bins are in $x_\gamma$ instead of $x_l$ 
are also shown. 
The EPA  elastic cross section agrees within $ 1 \% $ with 
the exact one for  all the $x_l$ bins. The agreement becomes slightly 
better if we consider $x_{\gamma}$ bins. For the inelastic channel, the
discrepancies from the EPA results are considerably higher. Our 'exact'
results lie closer to the EPA compared to \cite{lend} in almost all the
bins. The result in $x_{\gamma}$ bins show better agreement with the EPA
compared to the $x_l$ bins, especially for higher $x_\gamma$. The
discrepancy with the EPA is about $20-30 \%$ in most of the bins, higher in
some cases.

\section{summary}

In this work, we have calculated both elastic and inelastic QED Compton
scattering cross section in the unpolarized case. Our approach for the total
cross section is manifestly covariant and we have used the same cuts as in
the HERA-H1 experiment. The numerical estimates of the exact cross section for
different kinematical bins are presented and compared with the EPA
and the experimental results. The exact cross section in
the elastic channel agrees within $1 \%$ with the approximate one. The
discrepancy is thus due to the inelastic channel. The discrepancy with the
Monte Carlo estimate of \cite{lend} is also shown.
For both elastic and inelastic cross sections, our exact result is  closer 
to the EPA as compared to \cite{lend}. The agreement is even better if the
bins are in $x_{\gamma}$ instead of $x_l$. 
Our approach can be extended to calculate the 
corresponding cross section for polarized
scattering and also for other polarized and unpolarized processes having
photon induced subprocesses, in order to check how accurately the cross
section is given by the equivalent photon approximation. Also this would
predict the kinematical cuts necessary for the extraction of $\gamma(x,Q^2)$
experimentally from these processes.
   
\section{acknowledgements}

We warmly acknowledge M. Gl\"uck and E. Reya for initiating  this study, as
well as for  many helpful discussions and suggestions. We also thank V. Lendermann, 
M. Stratmann and I. Schienbein for
helpful discussions and comments. This work has been supported in part by
the 'Bundesministerium f\"ur Bildung und Forschung', Berlin/Bonn. 
   
\appendix 
\section{Kinematics in the $ e-\gamma $ c. m. Frame  (Elastic)}
In this appendix, we discuss the kinematics of the QED Compton scattering in 
the c. m. frame of the outgoing $e-\gamma$ system. The 4-momenta of the particles are given by:

\noindent
incident electron: $l \equiv (E_e^*,\,0,\,0,\,-E_e^*)$,

\noindent
incident proton: $ P \equiv (E_p^*,\,P_p^* {\sin{
\theta_p^*}},\,0,\,P_p^* {\cos{\theta_p^*}})$,  ~~~~ where~~$P_p^*=\sqrt
{{E_p^*}^2-m^2}$,

\noindent
outgoing electron: $l' \equiv E'^* (1,\,\sin \theta^* 
\cos \varphi^*,\,\sin \theta^* \sin \varphi^*,\, \cos \theta^* )$,

\noindent
outgoing photon: $ k' \equiv E'^* (1,\,-\sin \theta^* \cos 
\varphi^*,\,-\sin \theta^* \sin \varphi^*,\, -\cos \theta^*)$.

\noindent
We have also:
\noindent
4-momentum of the virtual photon: $ k=(E_k^*,\,0,\,0,\,E_e^*)$ ,~~ with $k^2=t$
and
\be
E_k^*=\sqrt {{E_e^*}^2+t}.
\label{ek}
\ee
The overall momentum conservation allows us to write the 4-momentum of the
final  proton as 
\be
P'=l+P-l'-k'.
\ee
We introduce the following Lorentz invariants 
\be
\hat s= (l'+k')^2= 4 {{E'}^*}^2,
\ee
\be
\hat t= (l-l')^2=-2 {E_e^*} {E'}^* (1+\cos \theta^*),
\ee
\be
\hat u= (l-k')^2=-2 {E_e^*} {E'}^* (1-\cos \theta^*),
\ee
\be
S= (l+P)^2= m^2 + 2 E_p^* E_e^*+2 E_e^* P_p^* \cos \theta_p^*,
\ee   
\be
T= (P-l')^2=m^2-2 {E'}^*(E_p^*-P_p^* \sin \theta^* \sin \theta_p^*\cos 
\varphi^*-P_p^* 
\cos\theta^* \cos \theta_p^*),
\label{cmT}
\ee
\be
U= (P-k')^2=m^2-2 {E'}^*(E_p^*+P_p^* \sin \theta^* \sin \theta_p^* \cos 
\varphi^*+P_p^*\cos\theta^* \cos \theta_p^*).
\label{cmU}
\ee
In addition they satisfy:
\be
\hat s+\hat t+\hat u=t, ~~~~~~S+T+U=-t+3 m^2.
\ee     
Using the relations above, it is possible to write the energies of the
particles in the lab frame in terms of the integration variables $\hat s$,
 $\hat t$ , $t$ and the constant $S$:
\be
E_e^*={\hat s-t\over 2 \sqrt {\hat s}}, ~~~~~E_k^*={\hat s+t\over 2 \sqrt 
{\hat s}},
\ee
and
\be
E_p^*={S-m^2+t\over 2 \sqrt {\hat s}}, ~~~~~P_p^*= {\sqrt{(S-m^2+t)^2-4 
\hat s m^2}\over 2 \sqrt {\hat s}},
 ~~~~~E'^* = {{\sqrt{\hat{s}}}\over{2}}.
\label{cmenergy}
\ee
Similarly, for the angles we have
\be
\cos \theta^* = { t- \hat s -2\hat t\over \hat s-t},
\label{cmangle1}
\ee
\be
\cos \theta_p^*={{2\hat s \,(S-m^2) -(\hat s -t)(S-m^2 + t)}
\over (S-t) [(S+t -m^2)^2 -4\hat s m^2]^{1\over 2}}.
\label{cmangle2}
\ee
In particular Eqs. (\ref{cmT}) and (\ref{cmU}), through
Eqs. (\ref{cmenergy})-(\ref{cmangle2}), express $T$ 
and $U$ in terms of our integrations variables  $\hat s$, $t$,
$\hat t$, $\varphi^*$ and we have used them to relate the lab frame 
variables to the integration ones, as shown in the next appendix.

\section{Kinematics in the lab frame (Elastic)}
In this appendix, we give the kinematics of  the same process in
the lab frame. The 4 momenta of the particles are given by:

\noindent
incident electron: $ l \equiv (E_e,\,0,\,0,\,-E_e)$,

\noindent
incident proton: $ P \equiv (E_p,\,0,\,0,\,P_p)$,  ~~~~ where~~$P_p^*=\sqrt
{{E_p}^2-m^2}$,

\noindent
outgoing electron: $ l' \equiv E_e'(1,\,\sin \theta_e,\,0,\,\cos
\theta_e)$,

\noindent
outgoing photon: $ k' \equiv E_{\gamma}'(1,\,\sin \theta_
{\gamma} \cos\phi_{\gamma},\,\sin \theta_{\gamma} \sin \phi_{\gamma},\,
\cos\theta_{\gamma})$.

Here we have chosen the frame such that the outgoing electron has zero
azimuthal angle. 
The Lorentz invariants are:
\be
\hat s=(l'+k')^2=2 E_e' E_{\gamma}' (1-\sin \theta_e \sin \theta_
{\gamma} \cos \phi_{\gamma}-\cos\theta_e \cos \theta_{\gamma}),
\label{shatl}
\ee
\be
\hat t= (l-l')^2=-2 E_e E_e' (1+ \cos \theta_e),
\ee
\be
\hat u=(l-k')^2=-2 E_e E_{\gamma}' (1+\cos \theta_{\gamma}),
\ee
\be
S=(l+P)^2=m^2+2 E_e (E_p+P_p),
\ee
\be
T=(P-l')^2=m^2-2 E_e' (E_p-P_p \cos \theta_e),
\ee
\be
U=(P-k')^2=m^2-2 E_{\gamma}' (E_p-P_p \cos \theta_{\gamma}).
\label{ul}
\ee
The polar angles in the lab frame can be written in terms of the invariants
and the incident energies:
\be
\cos \theta_e={E_p ~\hat t -E_e ~(T-m^2)\over  P_p ~\hat t +E_e ~(T-m^2)}
\label{thetae},
\ee
\be
\cos \theta_{\gamma}={E_p~(t-\hat s-\hat t)-E_e~(U-m^2)\over P_p~
(t-\hat s -\hat t)+E_e~(U-m^2)}.
\ee
In the same way, for the energies of the final electron and photon we
have: 
\be
E_e'=-{~\hat t~ P_p +E_e ~(T-m^2)\over S-m^2},
\ee
\be
E_{\gamma}'={P_p~(\hat s-t+\hat t)-E_e~(U-m^2)\over S-m^2}.
\ee
The azimuthal angle of the outgoing photon is: 
\be
\cos\phi_{\gamma}={2 {E_e'} {E_{\gamma}'} (1-\cos\theta_e\cos\theta_
{\gamma})-\hat s\over 2 
{E_e'} {E_{\gamma}'}
\sin \theta_e \sin \theta_{\gamma} },
\label{phig}
\ee
which is related to the acoplanarity angle by the relation $\phi=|\,\pi-
\phi_{\gamma}\,|$. Using Eqs. (\ref{cmT}) and (\ref{cmU}) together with Eqs.
(\ref{cmenergy})-(\ref{cmangle2}), the formulae above for $\cos \theta_e$, 
$\cos \theta_{\gamma}$, ${E_e'}$ , ${E_{\gamma}'}$ and $\cos\phi_{\gamma}$ 
can be  expressed in terms of the 
invariants and the incident energies. Eqs. (\ref{thetae})-(\ref{phig}) are
needed to implement numerically the kinematical region under study, since 
they relate the lab variables (energies and angles) to the ones used for the 
integration. 

\section{Kinematics in the $e-\gamma$  c. m. Frame
(Inelastic)}

In this frame most of the expressions remain the same as given in
appendix A. The relations among the invariants are now:
\be
\hat s+\hat t+\hat u=-Q^2,~~~~~~~~~~~~ S+T+U=3 m^2+{Q^2\over x_B},
\label{inv}
\ee
where $Q^2=-t$ and $x_B$ can be written as
\be
x_B={Q^2\over Q^2+W^2-m^2}.
\ee
The only formulae which are different are the ones involving $E_p^*$ and
$\cos\theta_p^*$. So Eq. (\ref{cmenergy}) will be replaced by   
\be
E_p^*={S-Q^2-W^2\over 2 \sqrt {\hat s}}, ~~~~~P_p^*={\sqrt{(S-Q^2-W^2)^2 -
4 \hat s m^2}\over 2 \sqrt {\hat s}},
 ~~~~~E'^* = {{\sqrt{\hat{s}}}\over{2}},
\label{cmenergyin}
\ee
while for the angles:
\be
\cos \theta^* = -{ Q^2 + \hat s +2\hat t\over \hat s + Q^2},
\label{cmangle1in}
\ee
and
\be
{\cos\theta_p^*}= {2 \hat s\, (S-m^2)-(S-Q^2-W^2) (\hat s+Q^2)\over (\hat s+Q^2)
{[(S-Q^2-W^2)^2-4 \hat s m^2]}^{1\over 2}}.
\label{cmangle2in}
\ee
Eqs. (\ref{cmenergyin})-(\ref{cmangle2in}) reduce to the Eqs. 
(\ref{cmenergy})-(\ref{cmangle2}) of the elastic channel for $W = m$ 
and $Q^2 = -t$.

\section{Kinematics in the lab frame (Inelastic)}
The invariants in the case of inelastic scattering are the same as in the
elastic case: Eqs. (\ref{shatl})-(\ref{ul}). 
Eq. (\ref{inv}) describes the relation among them.
The expressions of $\cos \theta_e$, $\cos \theta_{\gamma}$,  
$E_e'$, $E_{\gamma}'$ and $\cos \phi_{\gamma}$, in terms of the integration 
variables $W^2$, $\hat s$, $Q^2$, $\hat t$, $\varphi^*$  are given by Eqs. 
(\ref{thetae})-(\ref{phig}) together with Eqs. (\ref{cmT}), (\ref{cmU}) 
as before, but now Eqs. (\ref{cmenergyin})-(\ref{cmangle2in}) will replace 
Eqs. (\ref{cmenergy})-(\ref{cmangle2}).


\newpage
\vspace*{5cm}
\begin{center}
\epsfig{figure= 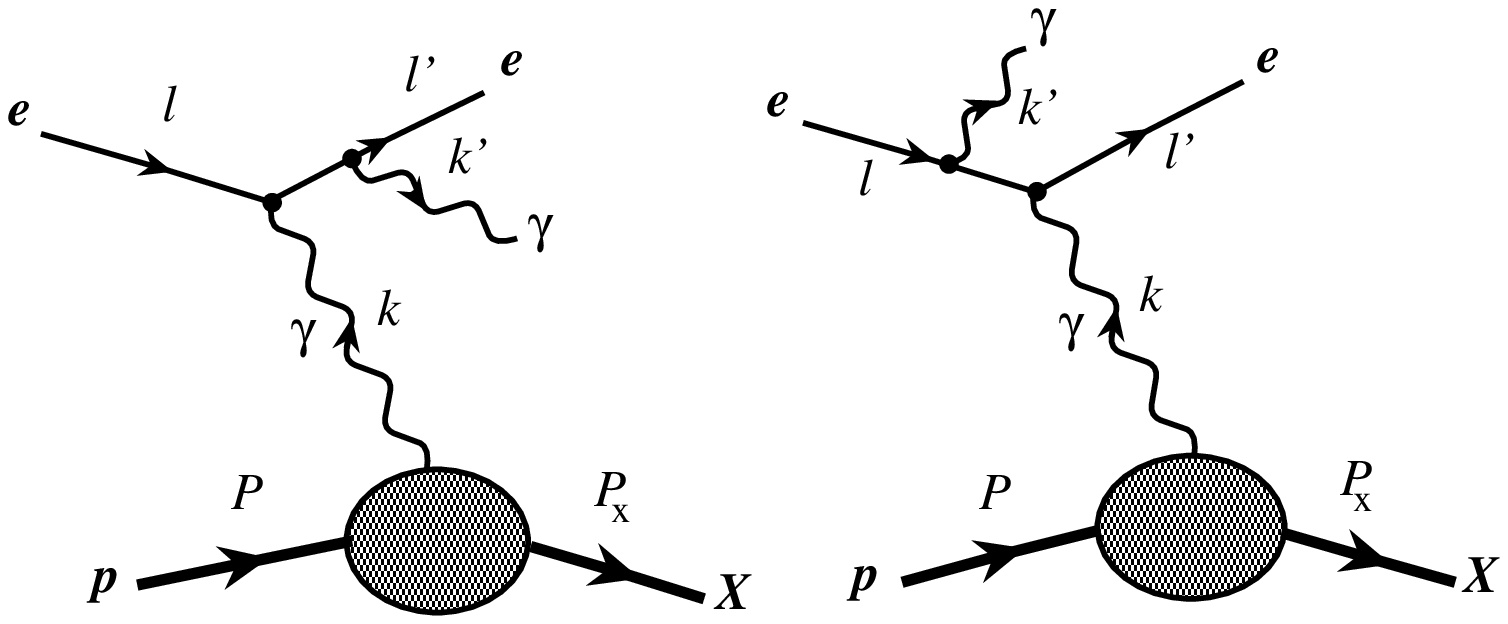, width=16cm, height= 8cm}\\
\end{center}
\begin{center}
\parbox{10.0cm}
{{\footnotesize 
 Fig. 1:  Feynman diagrams considered for $ep \rightarrow e\gamma X$, with \\
~~~~~~~~~~~ a real final state photon $(k'^2=0$).}}
\end{center}
\newpage
\begin{center}
\epsfig{figure= 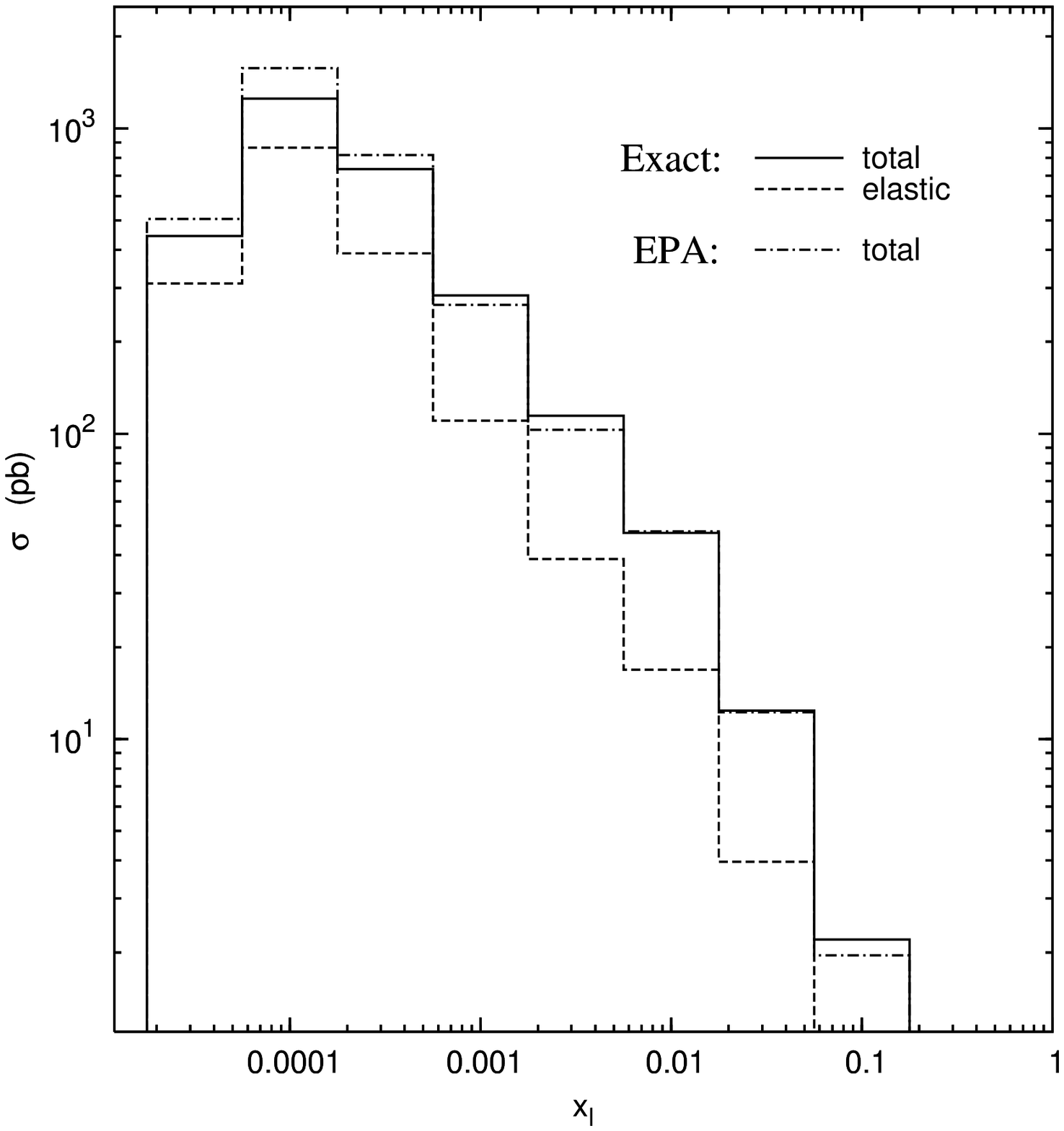,width=10.0cm,height=9.0cm}\\
\end{center}
\begin{center}
\parbox{14.0cm}
{{\footnotesize 
 Fig. 2:  Cross section for Compton process at HERA-H1. The cuts applied are as
described in the text.}}
\end{center}
\vspace{0.3cm}
\vspace{0.3cm}
\begin{center}
\epsfig{figure= 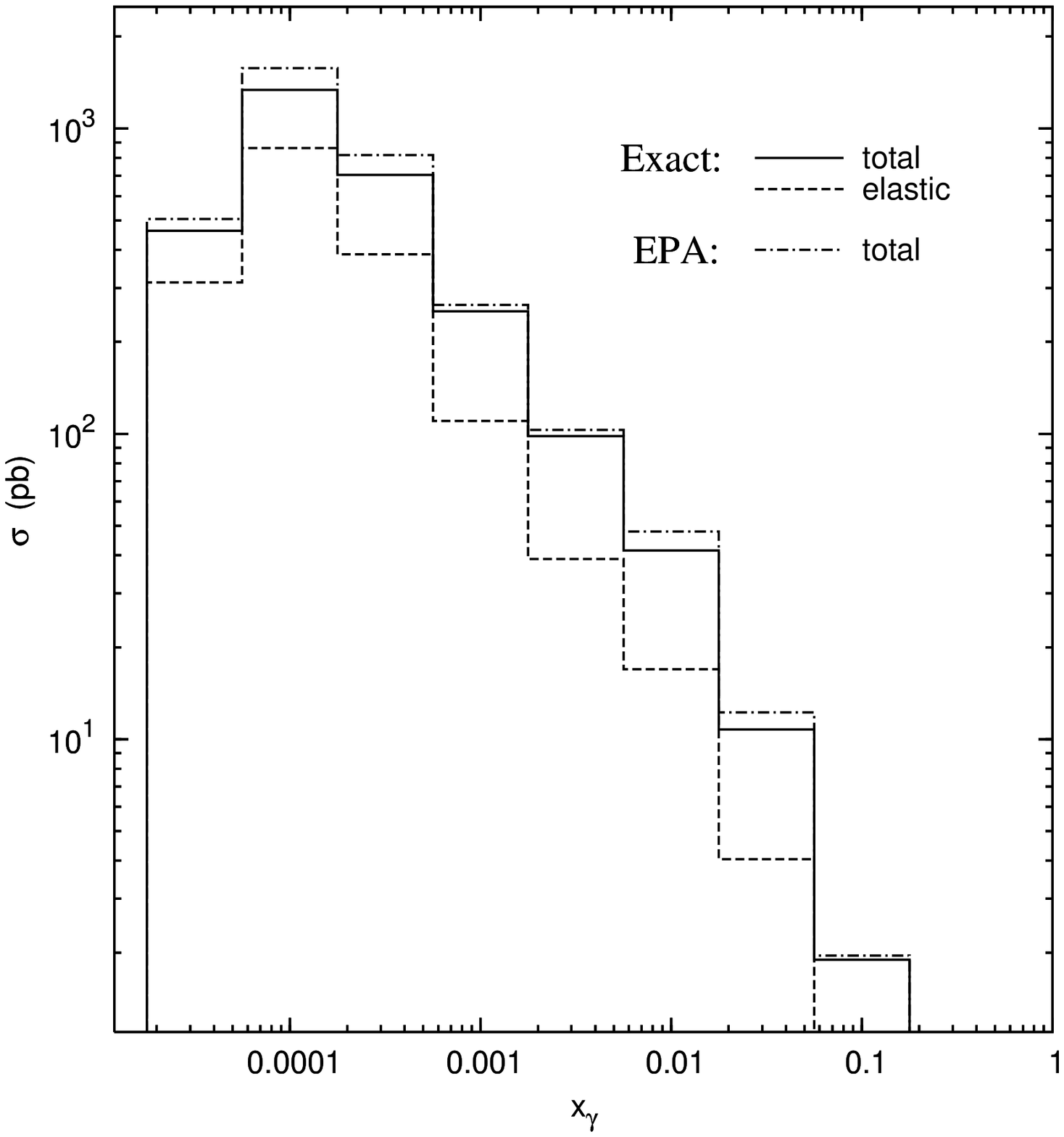,width=10.0cm,height=9.0cm}\\
\end{center}
\begin{center}
\parbox{14.0cm}
{{\footnotesize 
 Fig. 3:  Cross section for Compton process at HERA-H1. The bins are in
$x_{\gamma}$. The cuts applied are as
described in the text.}}
\end{center}
\newpage
\begin{center}
\parbox{8cm}{\epsfig{figure=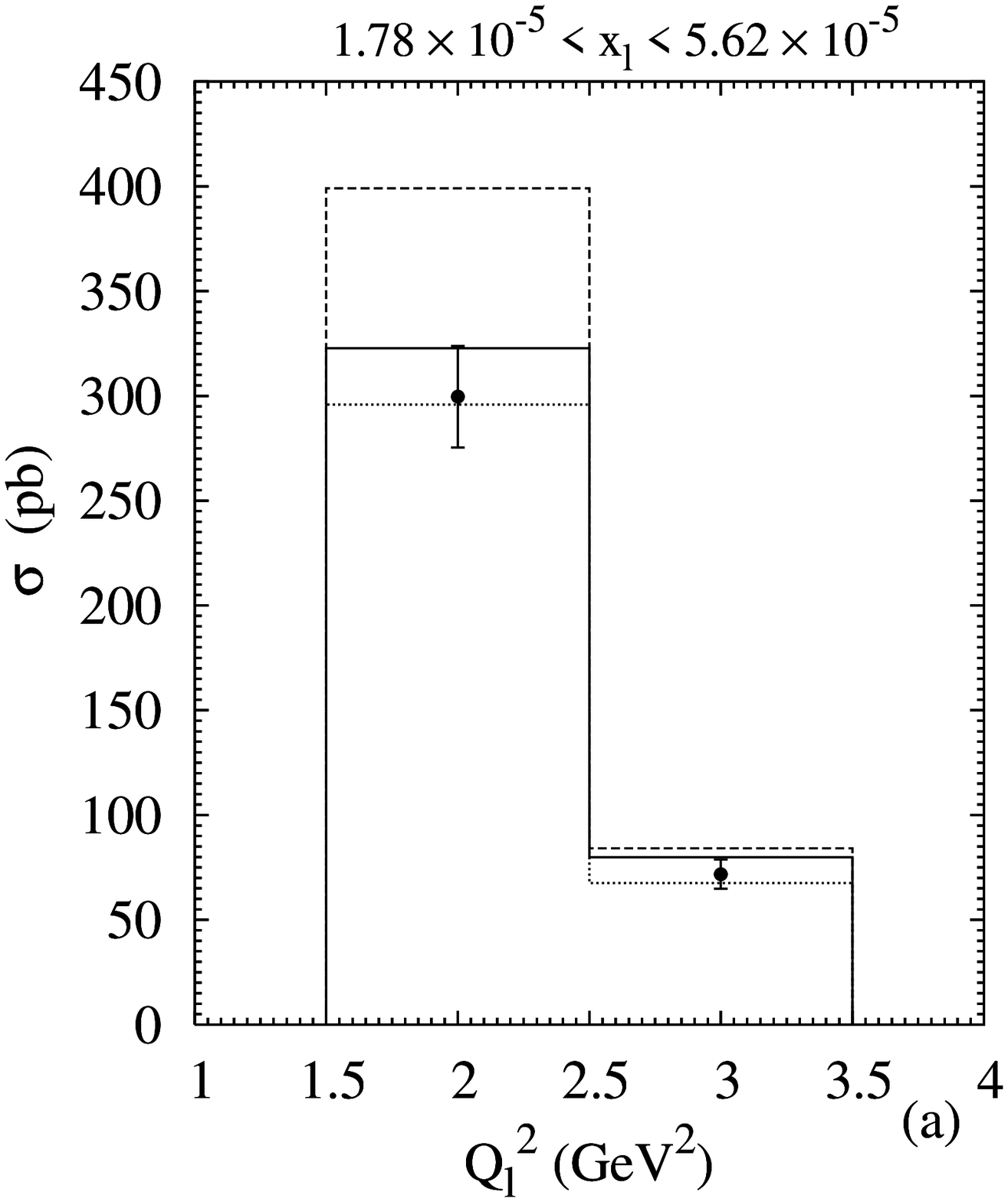,width=8.5 cm,height=7.5 cm}}\ \
\parbox{8cm}{\epsfig{figure=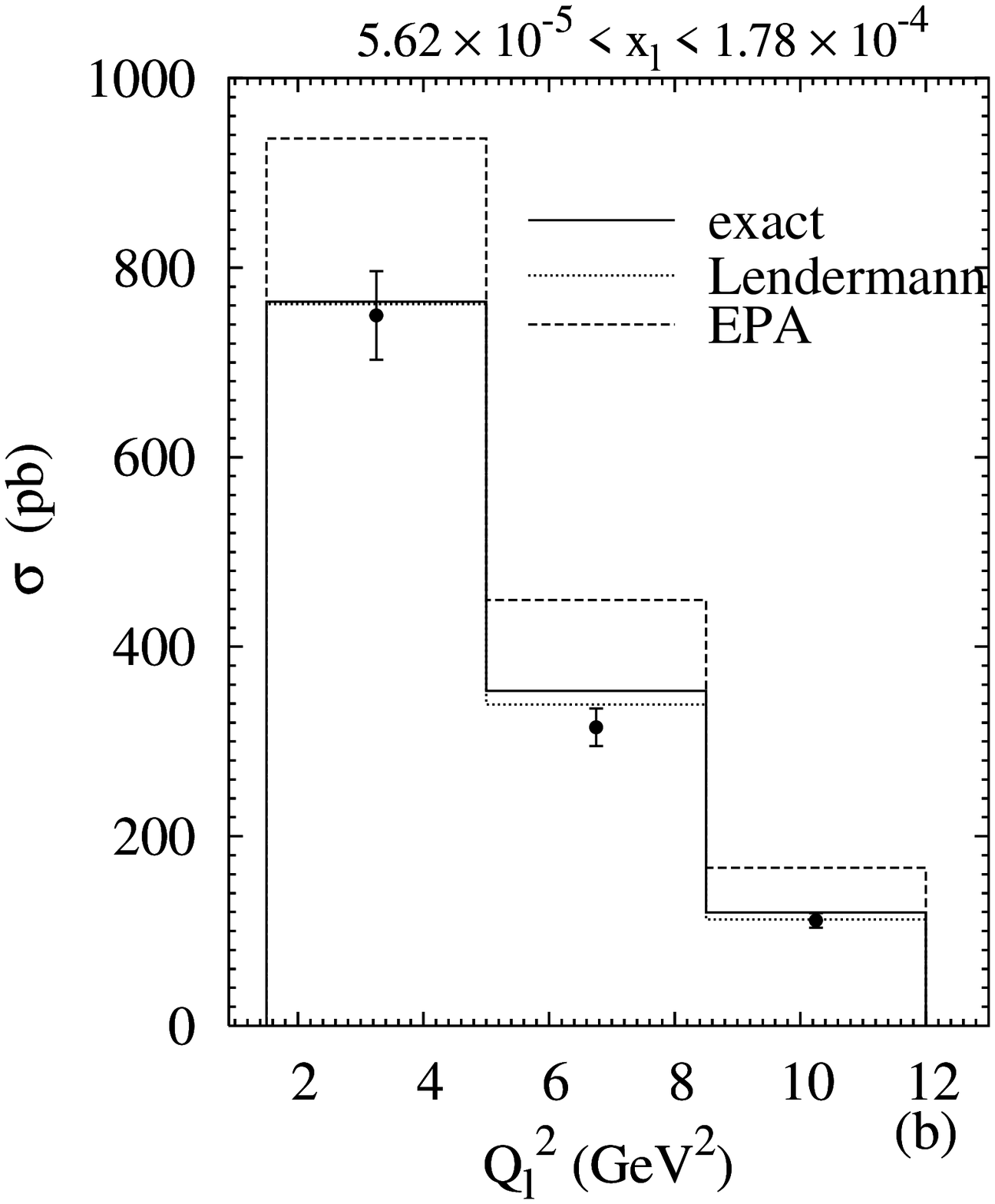,width=8.5 cm,height=7.5 cm}}\ \
\end{center}
\vspace{0.3cm}
\begin{center}
\hspace{-0.3cm}
\parbox{8cm}{\epsfig{figure=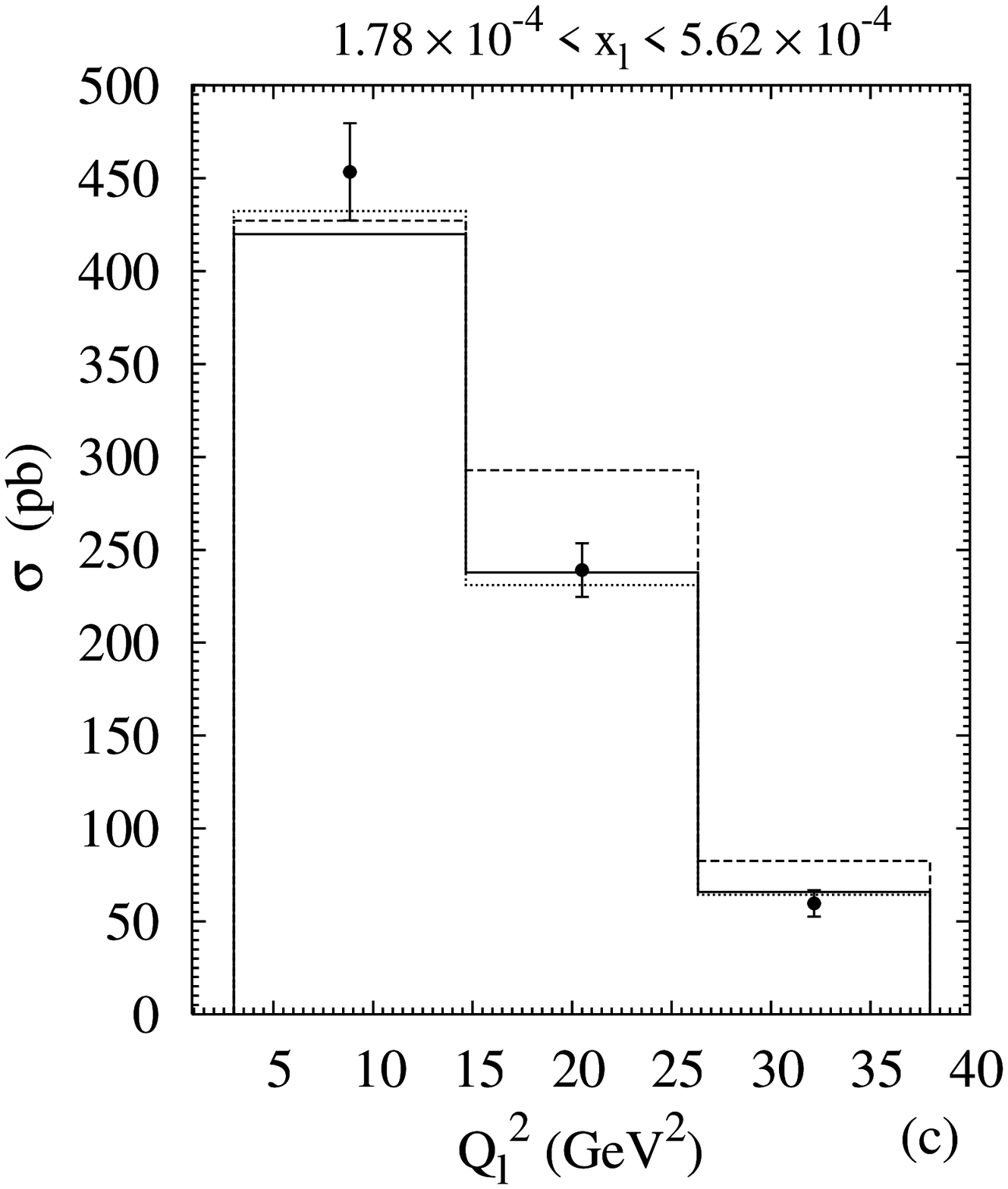,width=8.5 cm,height=7.5 cm}}\ \
\parbox{8cm}{\epsfig{figure=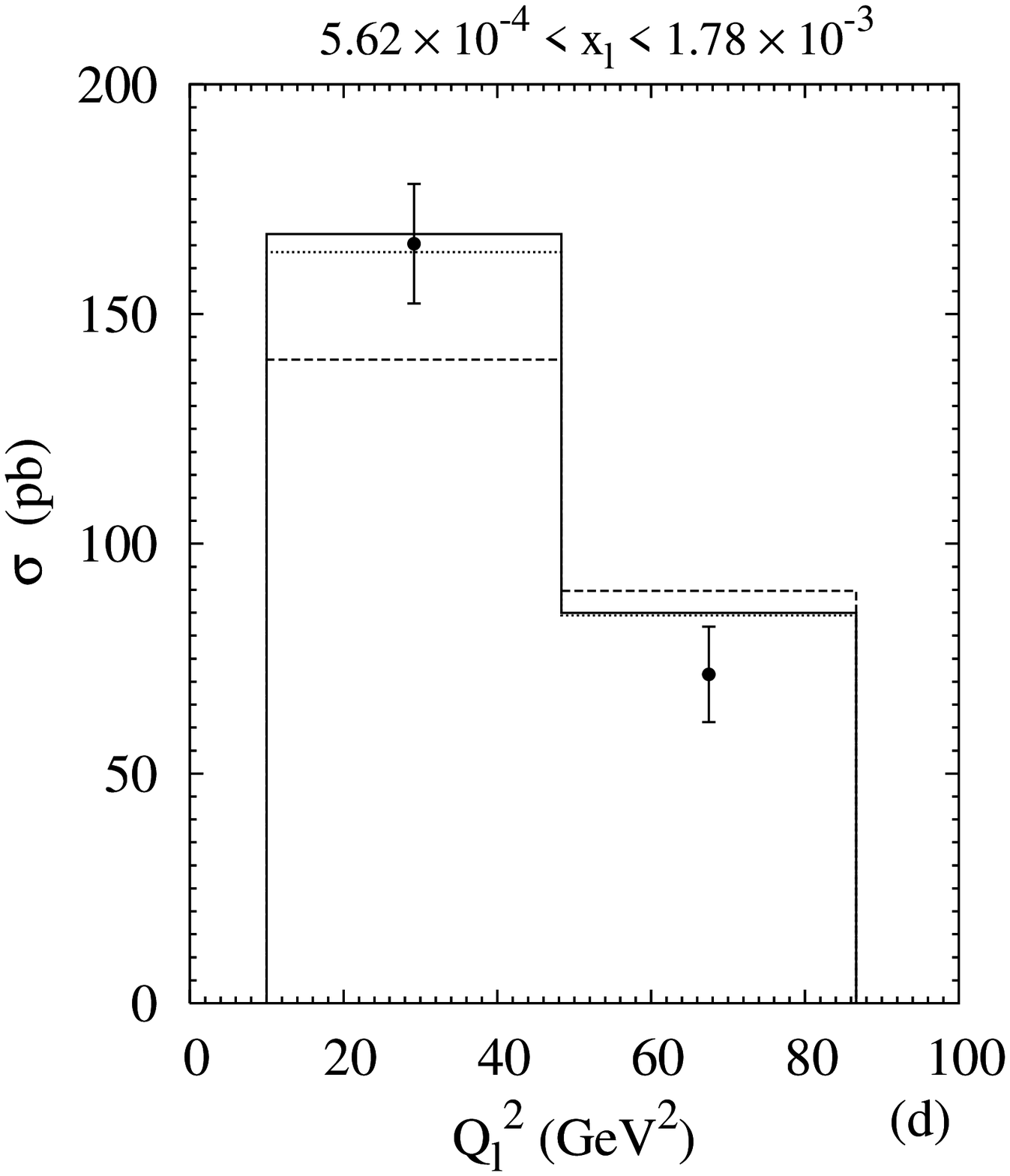,width=8.5 cm,height=7.5 cm}}\ \
\end{center}
\vspace{0.2cm}
\begin{center}
\parbox{14.0cm}
{{\footnotesize 
 Fig. 4:  Double differential cross section for QED Compton scattering at 
HERA-H1. The data are taken from \cite{lend}. The kinematical bins
correspond to Table 1. The continuous line corresponds to our exact
calculation, the dotted line to the calculation in \cite{lend} and the dashed line to the 
EPA.}}
\end{center}
\newpage
\begin{table}
\begin{center}
\begin{tabular}{|c|c|c|c|c|c|c|}
\hline 
$x$ bin & $Q^2_l$ bin & $\sigma_{\mathrm{el}}$  & 
$\sigma_{\mathrm{el}}^{\mathrm{Len}}$ &$\sigma_{\mathrm{el}}^*$ &
$\sigma_{\mathrm{el}}^{\mathrm{EPA}} $   \\ 
\hline \hline
          &           &             &     &      &  \\
$~~1.78\times 10^{-5}-5.62 \times 10^{-5}~~$ & $1.5 -2.5$ &$~~2.428\times 10^2~~$ & $2.342\times 10^2$   &$ 2.446\times 10^2  $ & $2.461 \times 10^2$    \\ 
$1.78\times 10^{-5}-5.62 \times 10^{-5}$ & $2.5 - 3.5$&$5.099\times 10^1$ & $~~4.71\times 10^1~~$ &$~~5.201\times 10^1~~$& $~~5.051\times 10^1~~$    \\ \hline
$5.62\times 10^{-5}-1.78\times 10^{-4} $ & $1.5 - 5.0$&$5.279\times 10^2$  &$5.319\times 10^2$ & $5.259\times 10^2$ & $5.247\times 10^2$ \\
$5.62\times 10^{-5}-1.78 \times 10^{-4}$&$ 5.0-8.5$&$2.396\times 10^2 $&$2.327\times 10^2$ & $2.404\times 10^2$ &  $2.395\times 10^2$ \\ 
$5.62 \times 10^{-5}-1.78\times 10^{-4}$ & $8.5-12.0$&$8.496\times 10^1 $ & $8.32\times 10^1$ & $8.559\times 10^1$ &$8.571\times 10^1 $ \\ \hline
$1.78\times 10^{-4}-5.62\times 10^{-4}$ & $3.0-14.67 $ & $2.080\times 10^2$ & $2.036\times 10^2 $ & $2.056\times 10^2$ &  $ 2.061\times 10^2$ \\ 
$1.78\times 10^{-4}-5.62\times 10^{-4}$ & $14.67-26.33 $ & $1.373\times 10^2$ & $1.388\times 10^2$ & $1.373 \times 10^2$ & $ 1.372\times 10^2$ \\ 
$1.78\times 10^{-4}-5.62\times 10^{-4}$ & $26.33-38.0 $ & $3.712\times 10^1$ & $3.86\times 10^1$ & $3.720 \times 10^1$   &$ 3.695\times 10^1$ \\ \hline
$5.62\times 10^{-4}-1.78\times 10^{-3}$ & $10.0-48.33 $ & $5.947\times 10^1$ & $5.71\times 10^1$ & $ 5.918\times 10^1$ & $ 5.921\times 10^1$ \\ 
$5.62\times 10^{-4}-1.78\times 10^{-3}$ & $48.33-86.67 $ & $3.714\times 10^1$ & $3.85\times 10^1$ &$3.715 \times 10^1$   & $ 3.704\times 10^1$ \\ 
$5.62\times 10^{-4}-1.78\times 10^{-3}$ & $86.67-125.0 $ & $1.056\times 10^1  $ & $1.028\times 10^1$ & $1.057\times 10^1$ & $1.054\times 10^1 $ \\ \hline
$1.78\times 10^{-3}-5.62\times 10^{-3}$ & $22-168 $ & $1.913\times 10^1$ & $1.877\times 10^1$ & $ 1.909\times 10^1 $ & $ 1.909\times 10^{1}$ \\ 
$1.78\times 10^{-3}-5.62\times 10^{-3}$ & $168-314 $ & $1.239\times 10^1$ & $1.229\times 10^1$ & $1.239 \times 10^1$& $ 1.238\times 10^1$ \\ 
$1.78\times 10^{-3}-5.62\times 10^{-3}$ & $314-460 $ & $5.917$ & $6.02$ & $5.915 $ & $ 5.914$ \\ \hline
$5.62\times 10^{-3}-1.78\times 10^{-2}$ & $0-500 $ & $4.811$ & $5.76$ & $4.890$ &  $4.890 $ \\ 
$5.62\times 10^{-3}-1.78\times 10^{-2}$ & $500-1000 $ & $9.271$ & $9.22$ & $9.264$ & $ 9.271$ \\ 
$5.62\times 10^{-3}-1.78\times 10^{-2}$ & $1000-1500  $ & $2.572 $ & $2.65$ & $2.571 $ & $ 2.573$ \\ \hline
$1.78\times 10^{-2}-5.62\times 10^{-2}$ & $0-1500 $ & $8.238\times 10^{-1}$ & $6.8\times 10^{-1}$ & $ 9.085\times 10^{-1}$ & $9.086\times 10^{-1}$ \\ 
$1.78\times 10^{-2}-5.62\times 10^{-2}$ & $1500-3000 $ & $2.431$ & $2.69$ & $2.430 $ &  $ 2.434$ \\ 
$1.78\times 10^{-2}-5.62\times 10^{-2}$ & $3000-4500  $ & $6.336\times 10^{-1}$ & $7.7\times 10^{-1}$ & $6.328 \times 10^{-1}$ & $6.345\times 10^{-1}$ \\ \hline
$5.62\times 10^{-2}-1.78\times 10^{-1}$ & $10-6005 $ & $3.120\times 10^{-1}$ & $4.27\times 10^{-1}$ & $3.120\times 10^{-1} $ & $ 3.117\times 10^{-1}$ \\ 
$5.62\times 10^{-2}-1.78\times 10^{-1}$ & $6005-12000 $ & $2.437\times 10^{-1}$ & $2.13\times 10^{-1}$ & $ 2.438\times 10^{-1} $ & $2.436\times 10^{-1} $ \\ 
$5.62\times 10^{-2}-1.78\times 10^{-1}$ & $~~12000-17995~~$ & $0.000$ &$0.000$ & $0.000$ & $2.461\times 10^{-2} $ \\ 
     &    &          &        &         &    \\
\hline

\end{tabular}
\end{center}
\caption{Double differential (elastic) QED Compton scattering cross section.
$\sigma_{\mathrm{el}}$ is the exact result in Eq. (\ref{sigel}), $\sigma_{\mathrm{el}}^{\mathrm{Len}}$ 
corresponds to the results in \cite{lend}. The $x$-bins refer to $x_l$ in
Eq. (\ref{xl}) except for $\sigma_{\mathrm{el}}^*$ where they refer to 
$x_\gamma$ in Eq. (\ref{xgamma}). $\sigma_{\mathrm{el}}^{\mathrm{EPA}}$ 
is given in Eq.
(\ref{epael}) where $x \equiv x_\gamma$. $Q^2_l$ is expressed in ${\mathrm{GeV}^2}$ and the cross-sections are in pb.}    
\label{tableone} 
\end{table}    

\newpage
\begin{table}
\begin{center}
\begin{tabular}{|c|c|c|c|c|c|c|}
\hline 
$x$ bin & $Q^2_l$ bin & $\sigma_{\mathrm{inel}}$  & 
$\sigma_{\mathrm{inel}}^{\mathrm{Len}}$ &$\sigma_{\mathrm{inel}}^*$ &
$\sigma_{\mathrm{inel}}^{\mathrm{EPA}} $    \\ 
\hline \hline
          &           &             &     &      &  \\
$~~1.78\times 10^{-5}-5.62 \times 10^{-5}~~$ & $1.5 -2.5$ &$~~7.996\times 10^1~~$ & $6.170\times 10^1$ & $ 7.503\times 10^1  $ & $1.529 \times 10^2$    \\ 
$1.78\times 10^{-5}-5.62 \times 10^{-5}$ & $2.5 - 3.5$&$2.880\times 10^1$ & $~~2.050\times 10^1~~$ &$~~4.142\times 10^1~~$& $~~3.361\times 10^1~~$    \\ \hline
$5.62\times 10^{-5}-1.78\times 10^{-4} $ & $1.5 - 5.0$&$2.361\times 10^2$  &$2.296\times 10^2$ & $2.364\times 10^2$ & $4.116\times 10^2$ \\
$5.62\times 10^{-5}-1.78 \times 10^{-4}$&$ 5.0-8.5$&$1.139\times 10^2 $&$1.062\times 10^2$ & $1.500\times 10^2$ &  $2.099\times 10^2$ \\ 
$5.62 \times 10^{-5}-1.78\times 10^{-4}$ & $8.5-12.0$&$3.442\times 10^1 $ & $2.890\times 10^1$ & $6.291\times 10^1$ &$8.094\times 10^1 $ \\ \hline
$1.78\times 10^{-4}-5.62\times 10^{-4}$ & $3.0-14.67 $ & $2.119\times 10^2$ & $2.287\times 10^2 $ & $1.278\times 10^2$ &  $ 2.210\times 10^2$ \\ 
$1.78\times 10^{-4}-5.62\times 10^{-4}$ & $14.67-26.33 $ & $1.005\times 10^2$ & $9.230\times 10^1$ & $1.220 \times 10^2$ & $ 1.556\times 10^2$ \\ 
$1.78\times 10^{-4}-5.62\times 10^{-4}$ & $26.33-38.0 $ & $2.868\times 10^1$ & $2.570\times 10^1$ & $4.682 \times 10^1$   &$ 4.558\times 10^1$ \\ \hline
$5.62\times 10^{-4}-1.78\times 10^{-3}$ & $10.0-48.33 $ & $1.079\times 10^2$ & $1.064\times 10^2$ & $ 5.383\times 10^1$ & $ 8.092\times 10^1$ \\ 
$5.62\times 10^{-4}-1.78\times 10^{-3}$ & $48.33-86.67 $ & $4.779\times 10^1$ & $4.590\times 10^1$ &$5.690 \times 10^1$   & $ 5.272\times 10^1$ \\ 
$5.62\times 10^{-4}-1.78\times 10^{-3}$ & $86.67-125.0 $ & $1.315\times 10^1  $ & $1.132\times 10^1$ & $2.050\times 10^1$ & $1.587\times 10^1 $ \\ \hline
$1.78\times 10^{-3}-5.62\times 10^{-3}$ & $22-168 $ & $4.758\times 10^1$ & $4.917\times 10^1$ & $ 2.426\times 10^1 $ & $ 3.080\times 10^1$ \\ 
$1.78\times 10^{-3}-5.62\times 10^{-3}$ & $168-314 $ & $2.010\times 10^1$ & $1.735\times 10^1$ & $2.350\times 10^1 $& $ 2.058\times 10^1$ \\ 
$1.78\times 10^{-3}-5.62\times 10^{-3}$ & $314-460 $ & $6.940$ & $5.760$ & $9.303 $ & $ 1.021\times 10^1$ \\ \hline
$5.62\times 10^{-3}-1.78\times 10^{-2}$ & $0-500 $ & $1.482\times 10^1$ & $1.432\times 10^1$ & $7.327$ &  $8.825 $ \\ 
$5.62\times 10^{-3}-1.78\times 10^{-2}$ & $500-1000 $ & $1.228\times 10^1$ & $9.890 $ & $1.318\times 10^1$ & $ 1.687\times 10^1$ \\ 
$5.62\times 10^{-3}-1.78\times 10^{-2}$ & $1000-1500  $ & $3.135 $ & $2.600$ & $3.756 $ & $ 4.885$ \\ \hline
$1.78\times 10^{-2}-5.62\times 10^{-2}$ & $0-1500 $ & $3.585$ & $2.500$ & $ 1.623$ & $1.811$ \\ 
$1.78\times 10^{-2}-5.62\times 10^{-2}$ & $1500-3000 $ & $3.778$ & $2.150$ & $3.866$ &  $ 4.867$ \\ 
$1.78\times 10^{-2}-5.62\times 10^{-2}$ & $3000-4500  $ & $9.675\times 10^{-1}$ & $6.600\times 10^{-1}$ & $1.088$ & $1.341$ \\ \hline
$5.62\times 10^{-2}-1.78\times 10^{-1}$ & $10-6005 $ & $1.093$ & $1.460\times 10^{-1}$ & $6.589\times 10^{-1} $ & $ 7.147\times 10^{-1}$ \\ 
$5.62\times 10^{-2}-1.78\times 10^{-1}$ & $6005-12000 $ & $5.496\times 10^{-1}$ & $2.110\times 10^{-1}$ & $ 5.782\times 10^{-1} $ & $5.922\times 10^{-1} $ \\ 
$5.62\times 10^{-2}-1.78\times 10^{-1}$ & $~~12000-17995~~$ & $6.164\times 10^{-2}$ &$4.300\times 10^{-2}$ & $0.000$ & $6.791\times 10^{-2} $ \\ 
     &    &          &        &         &    \\
\hline

\end{tabular}
\end{center}
\caption{Double differential (inelastic) QED Compton scattering cross section.
$\sigma_{\mathrm{inel}}$ is the exact result in Eq. (\ref{siin}),
$\sigma_{\mathrm{inel}}^{\mathrm{Len}}$ 
corresponds to the results in \cite{lend}. The $x$-bins are as in Table I,
{\it i.e.} refer to $x_l$ in
Eq. (\ref{xl}) except for $\sigma_{\mathrm{inel}}^*$ where they refer to $x_\gamma$ in
Eq. (\ref{xgamma}). $\sigma_{\mathrm{inel}}^{\mathrm{EPA}}$ is given in Eq.
(\ref{epain}) where $x \equiv x_\gamma$. $Q^2_l$ is expressed in ${\mathrm{GeV}^2}$ and the cross-sections are in pb.} 
\label{tabletwo} 
\end{table}

\end{document}